\documentclass[sigconf,table,nonacm]{acmart}

\usepackage{xcolor}
\usepackage[utf8]{inputenc}
\usepackage{amsmath}
\usepackage{graphicx} 
\usepackage{siunitx}
\usepackage{array}
\usepackage[frozencache]{minted}
\usepackage{soulutf8}
\usepackage{subcaption}
\usepackage{tikz}
\usetikzlibrary{decorations.pathreplacing}
\usepackage{float}
\usepackage{flushend}

\usepackage{enumitem}
\usepackage{balance}
\usepackage{hyperref}
\usepackage{url}
\usepackage{xspace}
\usepackage{tcolorbox}
\usepackage{multirow}
\usetikzlibrary{shapes}
\usetikzlibrary{patterns, fit}
\usepackage{adjustbox}
\usepackage{pifont}
\setminted[java]{
xleftmargin=15pt,
framesep=2mm,
linenos=true,
breaklines=true,
tabsize=2,
encoding=utf8,
frame=none,
fontsize=\scriptsize,
escapeinside=|| 
}

\newcommand{\highlight}[1]{\begin{tcolorbox}[leftrule=0mm,rightrule=0mm,toprule=0mm,bottomrule=0mm,left=0pt,right=0pt,top=0pt,bottom=0pt]

#1
\end{tcolorbox}
}

\newcolumntype{g}{>{\columncolor{gray!30}}l}
\newcolumntype{h}{>{\columncolor{gray!30}}c}

\newcommand{\difuzer}[0]{\textsc{Difuzer}\xspace}
\newcommand{\susi}[0]{\textsc{SuSi}\xspace}
\newcommand{\flowdroid}[0]{\textsc{FlowDroid}\xspace}
\newcommand{\ts}[0]{\textsc{TriggerScope}\xspace}
\newcommand{\hm}[0]{\textsc{HsoMiner}\xspace}
\newcommand{\az}[0]{\textsc{AndroZoo}\xspace}
\newcommand{\gp}[0]{Google Play\xspace}
\newcommand{\darkhazard}[0]{\textsc{HsoMiner}\xspace}
\newcommand{\db}[0]{\textsc{DataBomb}\xspace}

\newcommand{\appsWithTriggersPartial}[0]{\num{259}\xspace}
\newcommand{\appsWithTriggersPartialPercent}[0]{\num{2.59}}
\newcommand{\numberTriggersPartial}[0]{\num{2435}\xspace}
\newcommand{\numberTriggerPerAppPartial}[0]{\num{8.2}\xspace}
\newcommand{\flowBeforeADPartial}[0]{\num{14.60}\xspace}
\newcommand{\flowAfterADPartial}[0]{\num{0.24}\xspace}
\newcommand{\analysisTimePartial}[0]{\num{33.54}\xspace}
\newcommand{\numberLibFiltered}[0]{\num{19}\xspace}
\newcommand{\numberAppsTimeoutPartial}[0]{\num{24}\xspace}
\newcommand{\numberAppsTimeoutPartialPercent}[0]{\num{0.24}}
\newcommand{\appsWithTriggersHolistic}[0]{\num{339}\xspace}
\newcommand{\appsWithTriggersHolisticPercent}[0]{\num{3.39}}
\newcommand{\numberTriggersHolistic}[0]{\num{5575}\xspace}
\newcommand{\numberTriggerPerAppHolistic}[0]{\num{16.4}\xspace}
\newcommand{\flowBeforeADHolistic}[0]{\num{17.43}\xspace}
\newcommand{\flowAfterADHolistic}[0]{\num{0.56}\xspace}
\newcommand{\analysisTimeHolistic}[0]{\num{35.63}\xspace}
\newcommand{\difuzerPrecision}[0]{\num{99.02}\xspace}
\newcommand{\numberLogicBombsInManual}[0]{\num{30}\xspace}
\newcommand{\numberLogicBombsInManualPercent}[0]{\num{29.7}}
\newcommand{\numberAdditionalLogicBombs}[0]{\num{16}\xspace}
\newcommand{\numberLogicBombsInDataset}[0]{\num{46}\xspace}
\newcommand{\numberAppsManuallyAnalyzed}[0]{\num{102}\xspace}
\newcommand{\benignWithTriggersPartial}[0]{\num{354}\xspace}
\newcommand{\benignWithTriggersPartialPercent}[0]{\num{3.54}}

\newcommand{\appsRemovedFromGP}[0]{\num{8}\xspace}
\newcommand{\searchSpaceReduced}[0]{\num{81.9}}
\newcommand{\totalLibs}[0]{\num{5982}\xspace}

\begin{document}

\title{Difuzer: Uncovering Suspicious Hidden Sensitive Operations in Android Apps
} 

\author{Jordan Samhi$^1$, Li Li$^2$, Tegawendé F. Bissyandé$^1$, Jacques Klein$^1$}
\affiliation{%
  \institution{$^1$SnT, University of Luxembourg, Luxembourg, firstname.lastname@uni.lu \\ 
  $^2$Monash University, Australia, firstname.lastname@monash.edu}
  \country{}
}

\makeatletter
\def\@IEEEpubidpullup{6.5\baselineskip}
\newcommand{\ostar}{\mathbin{\mathpalette\make@circled\star}}
\newcommand{\make@circled}[2]{%
  \ooalign{$\m@th#1\smallbigcirc{#1}$\cr\hidewidth$\m@th#1#2$\hidewidth\cr}%
}
\newcommand{\smallbigcirc}[1]{%
  \vcenter{\hbox{\scalebox{0.97778}{$\m@th#1\bigcirc$}}}%
}
\let\runtitle\@title
\makeatother

\begin{abstract}
One prominent tactic used to keep malicious behavior from being detected during dynamic test campaigns is {\em logic bombs}, where malicious operations are triggered only when specific conditions are satisfied.
Defusing logic bombs remains an unsolved problem in the literature. In this work, we propose to investigate Suspicious Hidden Sensitive Operations (SHSOs) as a step towards triaging logic bombs. To that end, we develop a novel hybrid approach that combines static analysis and anomaly detection techniques to uncover SHSOs, which we predict as likely implementations of logic bombs.
Concretely, \difuzer identifies SHSO entry-points using an instrumentation engine and an inter-procedural data-flow analysis.
Then, it extracts trigger-specific features to characterize SHSOs and leverages One-Class SVM to implement an unsupervised learning model for detecting abnormal triggers.    

We evaluate our prototype and show that it yields a precision of 99.02\% to detect SHSOs among which 29.7\% are logic bombs.
\difuzer outperforms the state-of-the-art in revealing more logic bombs while yielding less false positives in about one order of magnitude less time.
All our artifacts are released to the community.
\end{abstract}

\maketitle

\section{Introduction}
\label{sec:introduction}
Security and privacy in Android have become paramount given its pervasive use in a wide range of user devices, be it handheld, at home, or in the office~\cite{idc_market_share}.  
Yet, regularly, new threats are discovered, even in the official \gp app store~\cite{zdnet}. 
Typically, thousands of apps are regularly flagged by antivirus engines: for the year 2020 alone, the \az~\cite{Allix:2016:ACM:2901739.2903508} repository has collected over \num{228000} apps, among which over \num{10000} apps are flagged by at least five antivirus engines hosted by VirusTotal.
Addressing the spread of malware in app markets is therefore a prime concern for researchers and practitioners. In the last decade, several approaches have been proposed in the literature to automate malware identification. These approaches explore static analysis techniques~\cite{7792435,doi:10.1155/2015/479174,7546513,papp2017towards,zhao2020automatic}, dynamic execution~\cite{10.1145/2592791.2592796,van2013dynamic,zheng2012smartdroid}, or a combination of both~\cite{10.1007/978-3-319-56991-8_51,8441295,brumley2008automatically}, as well as the use of machine-learning~\cite{6298824,6735264}. 

While the aforementioned techniques have been proven effective on benchmarks, attacks evolve rapidly with increasingly sophisticated evasion techniques. Typically, malware writers rely on code obfuscation~\cite{10.1007/978-3-030-01701-9_10} to bypass static analyses. To evade detection during dynamic analysis, attackers seek to hide malicious code behind triggering conditions. These are known as {\em logic bombs}, the triggering conditions of which being varied. For example, a logic bomb could execute malicious instructions only at a specific time that is not likely to be reached when market maintainers dynamically analyze the software before it is distributed. 

Logic bombs can be used for any malicious activity such as adware~\cite{erturk2012case}, trojan~\cite{pieterse2012android}, ransomware~\cite{yang2015automated}, spyware~\cite{saad2015android}, etc.~\cite{zhou2012dissecting}.
Furthermore, as the trigger and the malicious code are generally independent of the core application code, logic bombs can easily be added in legitimate apps and repackaged for distribution~\cite{10.1145/2484313.2484315,8653409,8029433,10.1007/978-3-319-47560-8_9}.
Therefore, detecting logic bombs is of great importance, especially in mobile devices that carry much personal information.
However, due to the undecidable nature of this detection problem in general~\cite{10.2307/1990888}, and the fact that dynamic analyses will likely fail to detect such behaviors~\cite{agrawal2012detecting}, analysts explore static-analysis based heuristic or machine learning approaches to detect logic bombs.

A logic bomb is characterized by the fact that it implements a hidden sensitive operation. 
Therefore, recent works addressing logic bombs have focused on the identification of  Hidden Sensitive Operations (HSOs) as a target~\cite{pan2017dark}.
However, not all HSOs are logic bombs. 
Indeed, an HSO may be neither {\bf intentional} nor {\bf malicious}, while logic bombs always are.
In this work, we propose to identify {\bf Suspicious HSOs} (SHSO) towards triaging logic bombs among HSOs.
Indeed, we consider that an SHSO is an HSO that is likely implementing a logic bomb.
Further note that, in this study, we do not attempt to address a binary classification problem of discriminating malware from benign apps (e.g., by using logic bombs as a key criteria of maliciousness).
Instead, our ambition is to improve the detection of logic bombs, which are considered sweet spots for targeting the understanding of malware's malicious behaviors. 
Indeed, while the literature proposes a variety of approaches for predicting Android apps' maliciousness (i.e., malware detection), the community still seeks to make significant breakthroughs in the location of malicious code parts. Detecting logic bombs thus provides an opportunity to locate and characterize malicious code implemented as hidden sensitive operations.

Recent literature on Android has already approached the problem of detecting sensitive behavior triggered only when certain conditions are met. Such triggers are referred hereafter as {\em sensitive triggers}. \ts~\cite{fratantonio2016triggerscope} was proposed as a static analysis tool to detect logic bombs: its analyses are based on heuristics and are thus limited to certain trigger types (i.e., time-related, location-related, and SMS-related triggers). \ts further relies on symbolic execution, which reduces its capacity to scale to massive datasets. Unlike \ts, \darkhazard~\cite{pan2017dark} leverages a supervised learning approach with engineered features to reveal sensitive triggers.  \darkhazard, however, does not specifically target malicious triggers: it flags up to \num{20}\% of apps, which makes it inefficient for isolating dangerous triggers in the wild; it also takes on average \num{13} min/app, which makes it challenging to exploit for large-scale experiments.

HSO triggering conditions are typically implemented by \emph{if statements}.
A given app code, however, may contain from hundreds to thousands of such conditional statements.
Therefore, a major challenge in the research around HSO is to reduce the search space for accurately spotting suspicious sensitive triggers.
Our core idea towards achieving this ambition is to model specific trigger characteristics to spot SHSOs.

In this work, we propose a novel approach to identify suspicious hidden sensitive operations where we rely on an unsupervised learning technique to perform anomaly detection.
We intend to detect suspicious triggers deviating from the normality of the myriads of conditional checks performed in typical apps.
To do so, we explore specific trigger/behavior features to guide our detection system towards enumerating truly suspicious triggers and thus refine the search space for uncovering logic bombs. 
We propose \difuzer, a novel hybrid approach that combines \ding{182} code instrumentation to insert particular statements required for taint analysis, \ding{183} inter-procedural static taint analysis to find suspicious sensitive triggers, and \ding{184} anomaly detection to reveal \emph{Suspicious Hidden Sensitive Operations} in Android apps.

While the literature includes work~\cite{pan2017dark} that proposed supervised learning techniques for detecting HSOs, \difuzer relies on unsupervised learning to spot  ``abnormal'' triggers. 
Moreover, towards ensuring that the model is accurate in the detection of suspicious HSOs, \difuzer leverages features that are specifically-engineered to capture semantic properties of maliciousness.

The main contributions of our work are as follows:
\begin{itemize}[leftmargin=*]
\setlength\itemsep{0em}
     \item We propose \difuzer, a novel approach to detect SHSOs in Android apps. \difuzer combines code instrumentation, static inter-procedural taint tracking and anomaly detection techniques.

    \item We evaluate \difuzer and show its ability to reveal SHSOs with a \num{99.02}\% precision in less than 35 seconds on average per app, outperforming previous approaches.
    
    \item We demonstrate that the trigger- and behavior-specific features of \difuzer are relevant for triaging logic bombs among HSOs:  \numberLogicBombsInManualPercent\% of detected SHSOs are indeed confirmed as logic bombs.
    
    \item We compare \difuzer against a state of the art logic bomb detector, \ts: \difuzer reveals more logic bombs than \ts while yielding less false positives.  
    
    \item We further applied \difuzer on a dataset of ``benign'' apps from Google Play. By analysing the yielded SHSOs, \difuzer  contributed to suspect \appsRemovedFromGP adware apps, which Google removed from \gp after we have pointed them out.
    
    \item We release the \difuzer prototype in open-source and further make available to the research community the first Android logic bomb dataset, called \db: \url{https://github.com/Trustworthy-Software/Difuzer}
\end{itemize}
\section{Background and Definitions}
\label{sec:background_definition}

In this section, we first introduce \emph{Taint Analysis} and \emph{Anomaly Detection}, two techniques used in our approach. Then, we carefully define important concepts and finally succinctly give the context for our study.

\textbf{Taint analysis:} 
Taint analysis is a dataflow analysis that follows the flow of specific values within a program.
A variable $V$ is tainted when it gets a value from specific functions called \emph{sources}.
The taint is propagated to other variables if they receive a derivation of the value in $V$.
If a tainted variable is used as a parameter of specific functions called \emph{sinks}, it means that during execution, the value derived from a \emph{source} can be used as a parameter of a \emph{sink}. 
In this paper's context, we rely on taint analysis to check if the conditional expression involves sensitive data value(s).

\textbf{Anomaly detection:}
When analyzing data of the same class, several items can significantly differ from the majority. They are called \emph{outliers} and can be viewed as abnormal.
There are numerous techniques in the state-of-the-art for achieving this outlier detection in sets of data~\cite{10.1145/1541880.1541882}.
This paper relies on \emph{One-Class Support Vector Machine} (OC-SVM)~\cite{doi:10.1162/089976601750264965}, an unsupervised learning algorithm that learns common behavior based on features extracted in an initial dataset.
Once the model is learned, a prediction is performed by checking whether a new sample features make it more or less abnormal w.r.t. the common model.
In this paper's context, an anomaly is computed by considering distances among vectors representing  \emph{triggers}, i.e., a condition along with the behavior triggered.

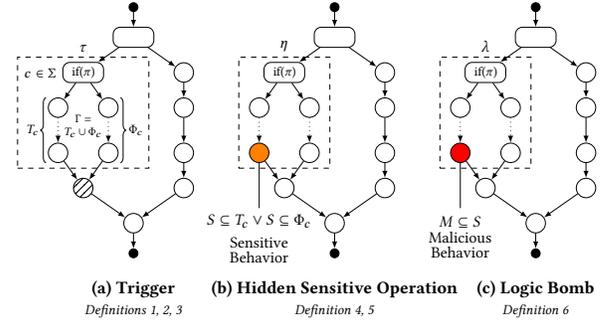
\begin{figure}[h]
    \centering
    \begin{adjustbox}{width=0.92\columnwidth,center}
    \begin{tikzpicture}
    \tikzstyle{arrowStyle}=[-latex]
    \tikzset{node/.style={minimum height=0.5cm, rounded corners, text width=.8cm,align=center,scale=.8}}
    
    \draw (0,-5.9) node[above]{\large \bf (a) Trigger};
    \draw (0,-6.3) node[above]{\small \textit{Definitions 1, 2, 3}};
    \draw (4,.-5.9) node[above]{\large \bf (b) Hidden Sensitive Operation};
    \draw (4,-6.3) node[above]{\small \textit{Definition 4, 5}};
    \draw (8,-5.9) node[above]{\large \bf (c) Logic Bomb};
    \draw (8,-6.3) node[above]{\small \textit{Definition 6}};

    \path
    (0,0) node[circle,draw=black, fill=black, inner sep=0pt,minimum size=5pt] (root) {}
    ++(0,-.6) node[node,draw] (a) {}
    ++(-1,-.7) node[node,draw] (b) {if($\pi$)}
    ++(2,0) node[circle,draw=black, fill=white, inner sep=0pt,minimum size=11pt] (c) {}
    ++(-2.5,-.7) node[circle,draw=black, fill=white, inner sep=0pt,minimum size=11pt] (d) {}
    ++(1,0) node[circle,draw=black, fill=white, inner sep=0pt,minimum size=11pt] (e) {}
    ++(1.5,0) node[circle,draw=black, fill=white, inner sep=0pt,minimum size=11pt] (f) {}
    ++(-2.5,-.9) node[circle,draw=black, fill=white, inner sep=0pt,minimum size=11pt] (g) {}
    ++(1,0) node[circle,draw=black, fill=white, inner sep=0pt,minimum size=11pt] (h) {}
    ++(1.5,0) node[circle,draw=black, fill=white, inner sep=0pt,minimum size=11pt] (i) {}
    ++(-2,-.7) node[circle,draw=black, fill=white, pattern=north east lines, inner sep=0pt,minimum size=11pt] (j) {}
    ++(2,0) node[circle,draw=black, fill=white, inner sep=0pt,minimum size=11pt] (k) {}
    ++(-1,-.7) node[circle,draw=black, fill=white, inner sep=0pt,minimum size=11pt] (l) {}
    ++(0,-.6) node[circle,draw=black, fill=black, inner sep=0pt,minimum size=5] (end) {}
    ;

  \draw[->,>=latex] (root) -- (a);
    \draw[->,>=latex] (a) -- (b);
    \draw[->,>=latex] (a) -- (c);
    \draw[->,>=latex] (b) -- (d);
    \draw[->,>=latex] (b) -- (e);
    \draw[->,>=latex] (c) -- (f);
    \draw[->,>=latex,dotted] (d) -- (g);
    \draw[->,>=latex,dotted] (e) -- (h);
    \draw[->,>=latex] (f) -- (i);
    \draw[->,>=latex] (g) -- (j);
    \draw[->,>=latex] (h) -- (j);
    \draw[->,>=latex] (i) -- (k);
    \draw[->,>=latex] (j) -- (l);
    \draw[->,>=latex] (k) -- (l);
    \draw[->,>=latex] (l) -- (end);

    \node[fit={(-2.2,-1.1) (-2.2,-3.1) (0.2,-1.1) (0.2,-3.1)}, draw, dashed] (t1) {};
    \draw (-1.85,-1.52) node[above]{\small $c \in \Sigma$};
    \draw (-1,-2.4) node[above]{\scriptsize $\Gamma =$};
    \draw (-1,-2.7) node[above]{\scriptsize $T_c \cup \Phi_c$};
    \draw (-1,-1) node[above]{$\tau$};

    \draw[decorate,decoration={brace,amplitude=3pt,raise=4pt},yshift=0pt] (-1.6,-3.1) -- (-1.6,-1.8) node[black,midway,xshift=-.4cm] {\footnotesize $T_c$};
    \draw[decorate,decoration={brace,amplitude=3pt,mirror,raise=4pt},yshift=0pt] (-.4,-3.1) -- (-.4,-1.8) node[black,midway,xshift=.45cm] {\footnotesize $\Phi_c$};

    \path
    (4,0) node[circle,draw=black, fill=black, inner sep=0pt,minimum size=5pt] (root1) {}
    ++(0,-.6) node[node,draw] (a1) {}
    ++(-1,-.7) node[node,draw] (b1) {if($\pi$)}
    ++(2,0) node[circle,draw=black, fill=white, inner sep=0pt,minimum size=11pt] (c1) {}
    ++(-2.5,-.7) node[circle,draw=black, fill=white, inner sep=0pt,minimum size=11pt] (d1) {}
    ++(1,0) node[circle,draw=black, fill=white, inner sep=0pt,minimum size=11pt] (e1) {}
    ++(1.5,0) node[circle,draw=black, fill=white, inner sep=0pt,minimum size=11pt] (f1) {}
    ++(-2.5,-.9) node[circle,draw=black, fill=orange, inner sep=0pt,minimum size=11pt] (g1) {}
    ++(1,0) node[circle,draw=black, fill=white, inner sep=0pt,minimum size=11pt] (h1) {}
    ++(1.5,0) node[circle,draw=black, fill=white, inner sep=0pt,minimum size=11pt] (i1) {}
    ++(-2,-.7) node[circle,draw=black, fill=white, inner sep=0pt,minimum size=11pt] (j1) {}
    ++(2,0) node[circle,draw=black, fill=white, inner sep=0pt,minimum size=11pt] (k1) {}
    ++(-1,-.7) node[circle,draw=black, fill=white, inner sep=0pt,minimum size=11pt] (l1) {}
    ++(0,-.6) node[circle,draw=black, fill=black, inner sep=0pt,minimum size=5] (end1) {}
    ;

    \draw[->,>=latex] (root1) -- (a1);
    \draw[->,>=latex] (a1) -- (b1);
    \draw[->,>=latex] (a1) -- (c1);
    \draw[->,>=latex] (b1) -- (d1);
    \draw[->,>=latex] (b1) -- (e1);
    \draw[->,>=latex] (c1) -- (f1);
    \draw[->,>=latex,dotted] (d1) -- (g1);
    \draw[->,>=latex,dotted] (e1) -- (h1);
    \draw[->,>=latex] (f1) -- (i1);
    \draw[->,>=latex] (g1) -- (j1);
    \draw[->,>=latex] (h1) -- (j1);
    \draw[->,>=latex] (i1) -- (k1);
    \draw[->,>=latex] (j1) -- (l1);
    \draw[->,>=latex] (k1) -- (l1);
    \draw[->,>=latex] (l1) -- (end1);
    
    \node[fit={(2.2,-1.1) (2.2,-3.1) (3.8,-1.1) (3.8,-3.1)}, draw, dashed] (t1) {};
    \draw (3,-1) node[above]{$\eta$};
    \draw (2.5,-4.5) node[above]{$S \subseteq T_c \vee S \subseteq \Phi_c$};
    \draw (2.5,-4.9) node[above]{Sensitive};
    \draw (2.5,-5.2) node[above]{Behavior};
    \draw[-,>=latex] (2.5,-4) -- (g1);

    \path
    (8,0) node[circle,draw=black, fill=black, inner sep=0pt,minimum size=5pt] (root2) {}
    ++(0,-.6) node[node,draw] (a2) {}
    ++(-1,-.7) node[node,draw] (b2) {if($\pi$)}
    ++(2,0) node[circle,draw=black, fill=white, inner sep=0pt,minimum size=11pt] (c2) {}
    ++(-2.5,-.7) node[circle,draw=black, fill=white, inner sep=0pt,minimum size=11pt] (d2) {}
    ++(1,0) node[circle,draw=black, fill=white, inner sep=0pt,minimum size=11pt] (e2) {}
    ++(1.5,0) node[circle,draw=black, fill=white, inner sep=0pt,minimum size=11pt] (f2) {}
    ++(-2.5,-.9) node[circle,draw=black, fill=red, inner sep=0pt,minimum size=11pt] (g2) {}
    ++(1,0) node[circle,draw=black, fill=white, inner sep=0pt,minimum size=11pt] (h2) {}
    ++(1.5,0) node[circle,draw=black, fill=white, inner sep=0pt,minimum size=11pt] (i2) {}
    ++(-2,-.7) node[circle,draw=black, fill=white, inner sep=0pt,minimum size=11pt] (j2) {}
    ++(2,0) node[circle,draw=black, fill=white, inner sep=0pt,minimum size=11pt] (k2) {}
    ++(-1,-.7) node[circle,draw=black, fill=white, inner sep=0pt,minimum size=11pt] (l2) {}
    ++(0,-.6) node[circle,draw=black, fill=black, inner sep=0pt,minimum size=5] (end2) {}
    ;
    
    \draw[->,>=latex] (root2) -- (a2);
    \draw[->,>=latex] (a2) -- (b2);
    \draw[->,>=latex] (a2) -- (c2);
    \draw[->,>=latex] (b2) -- (d2);
    \draw[->,>=latex] (b2) -- (e2);
    \draw[->,>=latex] (c2) -- (f2);
    \draw[->,>=latex,dotted] (d2) -- (g2);
    \draw[->,>=latex,dotted] (e2) -- (h2);
    \draw[->,>=latex] (f2) -- (i2);
    \draw[->,>=latex] (g2) -- (j2);
    \draw[->,>=latex] (h2) -- (j2);
    \draw[->,>=latex] (i2) -- (k2);
    \draw[->,>=latex] (j2) -- (l2);
    \draw[->,>=latex] (k2) -- (l2);
    \draw[->,>=latex] (l2) -- (end2);
    
    \node[fit={(6.2,-1.1) (6.2,-3.1) (7.8,-1.1) (7.8,-3.1)}, draw, dashed] (t1) {};
    \draw (7,-1) node[above]{$\lambda$};
    \draw (6.5,-4.5) node[above]{$M \subseteq S$};
    \draw (6.5,-4.8) node[above]{Malicious};
    \draw (6.5,-5.1) node[above]{Behavior};
    \draw[-,>=latex] (6.5,-4) -- (g2);
    
\end{tikzpicture}
    \end{adjustbox}
    \caption{Definitions illustrations. The graphs represent the Control-Flow Graph of the same function.}
    \label{fig:definitions}
\end{figure}

\textbf{Definitions:}
We define terms that will be used and referred to throughout the paper.
Figure~\ref{fig:definitions} visually depicts our definitions.

\noindent
\textbf{Definition 1} (Trigger).
A trigger is a piece of code that activates operations under certain conditions.
In Figure~\ref{fig:definitions}a, the trigger $\tau$ (dashed rectangle) is represented by the condition $c$ (rounded rectangle node), the true branch $T_c$ and the false branch $\Phi_c$.
The true branch $T_c$ represents all the statements (nodes) for which each path from the entry-point must go through $c$ and are executed if and only if $\pi$ is true. Note that every path from the entry-point to the hatched node must go through $c$. In other words, $c$ strictly dominates the hatched node. However, the hatched node can be executed if $\pi$ is true or false. Therefore it is not part of $T_c$ nor $\Phi_c$.
The false branch $\Phi_c$ represents all the statements for which each path from the entry-point must go through $c$ and are executed if and only if $\pi$ is false.

More formally, let $\Sigma$ be the set of statements of a function (nodes in Fig.~\ref{fig:definitions}).
Let $c \in \Sigma$ be a conditional statement (i.e., an if statement, rectangle nodes in Fig.~\ref{fig:definitions}).
Let $\pi$ be $c$'s predicate.
Let $\varepsilon$ be the conditional execution function such as $\varepsilon(\pi, \sigma)$ is true if $\sigma \in \Sigma$ is executed if and only if $\pi$ is true. 
Let $\delta$ be the dominator function such as $\delta(d,\sigma)$ is true if $d \in \Sigma$ strictly dominates $\sigma \in \Sigma$, false otherwise.

\noindent
Let $T_c$ and $\Phi_c$ be the \emph{true} and the \emph{false} branch~\footnote{Note that in case there is no false branch, $\Phi_c = \emptyset$.} of $c$ such as:

\begin{center}
$T_c = \lbrace \sigma \: \vert \: \sigma \in \Sigma \wedge \delta(c, \sigma) \wedge \varepsilon(\pi, \sigma)\rbrace$

$\Phi_c = \lbrace \sigma \: \vert \: \sigma \in \Sigma \wedge \delta(c, \sigma) \wedge \varepsilon(\neg \pi, \sigma)\rbrace$    
\end{center}

\noindent
Then, a trigger $\tau$ is defined as a triplet: $\tau = (c, T_c, \Phi_c)$.

\noindent
\textbf{Definition 2} (Guarded code). Let $\tau$ be a trigger such as: $\tau = (c, T_c, \Phi_c)$.
\noindent
Then, the code guarded by $c$ is: $\Gamma = T_c \cup \Phi_c$.

\noindent
\textbf{Definition 3} (Trigger entry-point). We define a trigger entry-point as the condition triggering the guarded code.
More formally, given a trigger $\tau = (c, T_c, \Phi_c)$, $c$ is defined as its entry-point.

\noindent
\textbf{Definition 4} (Hidden Sensitive Operation (HSO)). An HSO is a piece of code that represents a set of instructions, which (1) implement a security-sensitive operation and (2) are only executed when specific criteria are met (cf. Figure~\ref{fig:definitions}b).
More formally, let $\eta = (c, T_c, \Phi_c)$ be a trigger and $S$ a piece of sensitive behavior such as $S \subset \Sigma$.
Then, $\eta$ is a hidden sensitive operation if $S \subseteq T_c \vee S \subseteq \Phi_c$.

\noindent
\textbf{Definition 5} (Suspicious Hidden Sensitive Operation (SHSO)).
An SHSO refers to an HSO that implements a sensitive operation that appears to be suspicious given the context of the app. 
For example, a navigation app may legitimately retrieve user location information (which is a sensitive  operation), while a calculator is suspicious if it attempts to retrieve such sensitive data.

\noindent
\textbf{Definition 6} (Logic bomb). A logic bomb is a piece of malicious code triggered under specific circumstances.
More formally, let $\lambda = (c, T_c, \Phi_c)$ be an SHSO, $S$ its sensitive behavior, and $M$ a piece of malicious code such as $M \subset \Sigma$.
Then, $\lambda$ is a logic bomb if $M \subseteq S$ (cf. Figure~\ref{fig:definitions}c).
In other words, a logic bomb is an SHSO which suspicious sensitive behaviour is malicious.

\begin{listing}[h]
\begin{minted}{java}
// Example simplified for reading, with renamed methods
public static void m() {
    m1();
    performMaliciousActivity();
}
public static void m1() {
    if (m2()){System.exit(0);}
}
public static boolean m2() {
    String str1 = Build.MODEL;
    String str2 = encryptedString(); // str2 = "Emulator"
    return str1.contains(str2);
}
public static String encryptedString() {
    String s1 = "cb6624dec24f889f4fcdf6c8cda99d4a";
    return BYDecoder.decode(s1, "0ec47edd8db3a02b");
}
\end{minted}
    \caption{Logic bomb identified by \difuzer in  "com.flyingbees.BrasilTvEnvivo" with emulator evasion.}
    \label{code:emulator}
\end{listing}

In Listing~\ref{code:emulator}, we summarize the general behavior of a concrete example of a logic bomb extracted from a real-world app. This logic bomb was detected by \difuzer.
In this example, the different parts of the SHSO (including the triggering condition checks) are split across several methods  ($m1, m2, m$). The actual triggering condition check is done in line 3: $m2$ will return {\tt true} if the device runs in an emulator and the app execution will be halted. Otherwise, the malicious behavior (line 2) will be triggered.

The challenge in detecting the aforementioned logic bomb is that analysts cannot rely on rules or models to detect it due to the lack of a formal definition of malicious behavior. Therefore, we note that, with little coding effort,  malware authors could push malicious code that will be missed in most dynamic analyses.
Indeed, sandboxes and testing environments usually return default values for environment variables~\cite{10.1145/2592791.2592796}.
Besides the device's model, different environment values (e.g., sensors, settings, GPS, remote values, etc.) can be used to trigger malicious code.

\begin{tcolorbox}[leftrule=0mm,rightrule=0mm,toprule=0mm,bottomrule=0mm,left=0pt,right=0pt,top=0pt,bottom=0pt]
Comparing to previous works, we note that the presented simple example of logic bomb detected by \difuzer would constitute a challenge to the existing state of the art.  \ts~\cite{fratantonio2016triggerscope} cannot identify this logic bomb. Indeed, since its heuristics are limited to time-, location-, and SMS-related triggers, logic bombs with a new trigger (e.g., environment variable such as \texttt{Build} class fields) are missed.
\darkhazard~\cite{pan2017dark} could detect this logic bomb if its training set includes similar examples. Unfortunately, \darkhazard flags too many HSOs (e.g.,  $\sim$\num{20}\% of apps), making the manual check a cumbersome task. 
In contrast, \difuzer offers a reasonable number of warnings to be checked manually.
\end{tcolorbox}
\vspace{-.5cm}
\section{Approach}
\label{sec:approach}

\begin{figure*}
    \centering
    \begin{adjustbox}{width=.8\linewidth,center}
    \input{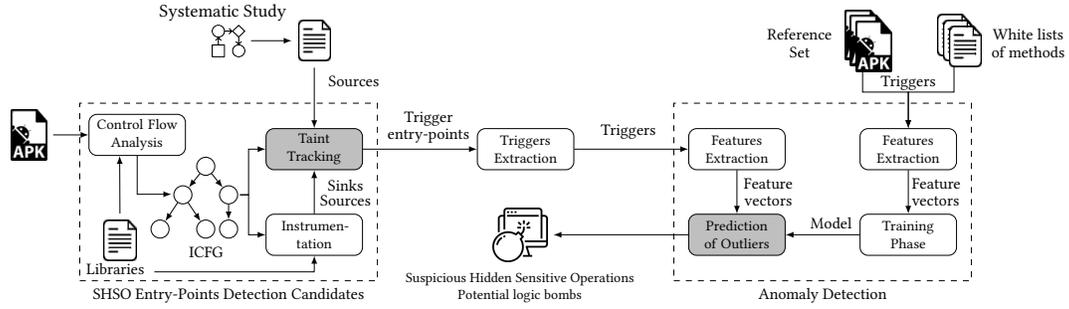}
    \end{adjustbox}
    \caption{Overview of the \difuzer approach.}
    \label{fig:overview}
\end{figure*}

\noindent
\textbf{Goal:} 
With \difuzer, we do not aim at detecting any HSOs, but only suspicious HSOs (SHSOs) for which the likelihood of being logic bombs is high.

\noindent
\textbf{Intuition:} 
As shown in previous studies~\cite{pan2017dark}, the number of HSOs per app can be large, even in benign apps. This suggests that although HSOs are "sensitive" operations, most of them are legitimate, i.e., they are used to implement common behavior. 
In contrast, logic bombs are rare, especially in benign apps. The idea behind \difuzer is to use an anomaly detection approach, with specifically designed features, to triage logic bombs among SHSOs.

\noindent
\textbf{Overview:} In Figure~\ref{fig:overview}, we present an overview of our approach, which consists of two main modules:
(1) identification of SHSO entry-point candidates via control flow analysis, instrumentation, and taint tracking (left dotted block); 
(2) From these entry-points, triggers are extracted, and the second module (right dotted block) extracts specifically designed features fed into an outliers predictor. This predictor is previously trained on a set of reference apps (i.e., apps considered benign) to learn legitimate usages of triggers.

\subsection{Identifying SHSO candidate entry-points}
\label{sec:identify_entrypoints}

Previous works~\cite{evasion_techniques,tampering_detection,android_hehe,hacking_team,10.1145/2592791.2592796} have shown that specific values, such as system inputs and environments variables, are often used to trigger HSOs.
State-of-the-art approaches have thus proposed to check whether the conditions of \emph{if statements} contain these sensitive data.
To that end, they rely on symbolic execution~\cite{fratantonio2016triggerscope} or backward data-dependency graphs~\cite{pan2017dark} that could suffer from scalability problems.
With \difuzer, we propose to use taint analysis to track sensitive data values and check if they are involved in conditional expressions.

Taint analysis tools generally track data from sources to sinks.
The implementation of \flowdroid, a popular taint analysis framework for tracking sensitive information, considers sources and sinks at the method level.
In our case however, sinks are fine-grained code locations, which are conditional expressions of \emph{if statements}. This requires for \difuzer to instrument apps in order to insert dummy method calls that will make the apps ready for analysis by \flowdroid (cf. Section~\ref{sec:instrumentation}).
Moreover, sources can be method calls or data field accesses. To build the set of source and sinks we propose to make a systematic mapping (cf. Section~\ref{sec:sources_sinks})  that explores internal and external system properties and their associated APIs as well as environment variables.

\subsubsection{Systematic mapping toward defining sources}
\label{sec:sources_sinks}
As already explained, a first step is to track sensitive values. 
In this work, these values are derived from particular source methods. 
Then, if a sensitive value falls into an \emph{if statement}, we consider the condition as a potential SHSO entry-point.
This section will describe how we gathered a comprehensive list of source methods used for the taint tracking phase.
Note that we did not rely on the reference sources list produced by \susi~\cite{arzt2013susi} since it has been shown that most of the methods are inappropriate for tracking sensitive data, and lead to a high amount of false-positives (e.g., $>$80\%)~\cite{8952502,nan2018finding,junaid2016dexteroid}.

In general, decisions on whether to trigger SHSOs or not are taken on system properties~\cite{evasion_techniques,android_hehe,path_to_payload,pan2017dark}. Hence, we performed a systematic mapping of the Android framework from SDK version 3 to 30 (versions 1 and 2 were unavailable) to gather a comprehensive list of source methods.
In particular, since in the case of Android apps, system properties can be derived from the device's \emph{internal} and \emph{external} properties, we inspect the successive versions of the framework to identify various means to access these properties. 

\begin{table}[ht!]
      \begin{adjustbox}{width=\columnwidth,center}
          \begin{tabular}{|l|c|c|c|c|c|c|}
            \hline
            \multirow{3}{*}{} & \multicolumn{6}{c|}{\textbf{Device}} \\
            \cline{2-7}
            & \multicolumn{3}{c|}{\textbf{Internal}} & \multicolumn{3}{c|}{\textbf{External}} \\
            \cline{2-7}
            & \textbf{System} & \textbf{Content} & \textbf{Build} & \textbf{SIM} & \textbf{Internet} & \textbf{GPS} \\
            \hline
            \multirow{2}{*}{Examples} & Sensors, & Call Logs, & Model, & Phone call, & Parameters, & Latitude, \\
            & Camera & Contacts & Hardware & SMS & Content & Longitude \\
            \hline
            \end{tabular}
        \end{adjustbox}
    \caption{Examples of sensitive sources}
    \label{table:sensitive_sources}
\end{table}

In Table~\ref{table:sensitive_sources}, we enumerate the different property types (with examples) on which we reasoned to retrieve sensitive sources, which are classically focused on in the literature~\cite{evasion_techniques,android_hehe,path_to_payload,pan2017dark}.
We follow a systematic process to perform the retrieval of sources from the given property types: we first extracted patterns from the different ways to access the aforementioned properties.
Then, we used those patterns to automatically discover the sensitive sources that we make available to the research community in the \difuzer project's repository.
In the following, we further detail the internal and external properties that we consider.

\noindent
\textbf{Internal:} In the case of internal properties, a developer can get sensitive information of the device from three main channels: 1) System properties, 2) Content in internal databases, and 3) Information from \texttt{BUILD} class (see Table~\ref{table:sensitive_sources}). In the following, we describe how we obtain a list of sources for those three channels:

\noindent
\ding{182} \textit{System properties}: While developing an Android app, developers have access to several useful APIs. In this case, the most interesting is \texttt{android.content.Context.get\-System\-Ser\-vi\-ce\-(java.\-lang.\-String)}~\cite{context} which returns the system-level handler for a given service. The service is described by a string given as parameter to \texttt{get\-System\-Service} method. The \texttt{Context} class gives developers access to pre-defined constants (e.g., \texttt{SENSOR\_SERVICE}). 

In fact, every constant contains the name of the service with \texttt{"\_SERVICE"} appended to it. The return value type of the \texttt{getSystem\-Service} method call is derived from the constant name (e.g., \texttt{SENSOR\-SERVICE} will give a \texttt{Sensor\-Manager}~\cite{sensormanager}) which in turn can be used to get a object whose type is also derived from the constant name (e.g., a Sensor\-Manager object can be used to obtain a Sensor object~\cite{sensor}).
We used this pattern to compile our list of sensitive sources for the System properties. More specifically, we verify if the class exists in at least one SDK version for each class obtained.
If this is the case, we list the methods of the class and keep only the "getter methods",  i.e., those starting by "get" or "is" (e.g., methods such as \texttt{getId()} or \texttt{isWifiEnabled()}). 

\noindent
\ding{183} \textit{Content in internal databases}: To access databases fields, one has to perform a query which returns a \texttt{android.database.Cursor}~\cite{cursor} object. This object is then used to iterate over the result of the query. Hence, to get sensitive source methods related to content in internal databases,  
we applied the same process as for system properties (i.e., to retrieve the "getter" methods) but on the \texttt{Cursor} class.

\noindent
\ding{184} \textit{\texttt{Build} class}: The \texttt{Build} class~\cite{build} allows developers to access information about the current build of the device from its fields. For instance, one can get the brand associated with the device by accessing \texttt{Build.BRAND}. 
Note that our objective is to retrieve a list of source methods. However, the information a developer can get from the \texttt{Build} class can only be retrieved from class fields, not method calls. 
Consequently, in  Section~\ref{sec:instrumentation}, we will explain how we instrument the app under analysis to add method call statements representing \texttt{Build} field accesses.

We gathered a list of 618 unique methods for internal values.

\noindent
\textbf{External:} In the case of external properties, a developer can get sensitive information from three channels: 1) SIM card, 2) Internet Connection, and 3) GPS chip. 
The process to collect the source methods is similar to the one followed with \texttt{Cursor} class, except we do not know in advance the name of the classes to inspect. Therefore we relied on a heuristic to identify such classes: for each SDK version, we listed all the classes and kept only those with class names containing the following words: "Sms, Telephony, Location, Gps, Internet, and Http". 
Once the classes were retrieved, we listed the methods for each class and kept those starting by "get" or "is". The intuition is the same as in the case of internal sources.

We gathered a list of 794 unique methods for external values.
Finally, after combining sensitive sources from internal and external values, our list contains 1285 unique methods (127 duplicates).

\subsubsection{Instrumentation}
\label{sec:instrumentation}

Performing taint tracking, as briefly described in Section~\ref{sec:background_definition}, consists of a data-flow algorithm that propagates the taint from a source method to a sink method.

\noindent
\textbf{Sinks related challenge:} 
We remind that one objective of \difuzer is to identify SHSOs' trigger entry-points. 
Consequently, the taints that \difuzer tracks are supposed to fall into \emph{if statements}. 
However, being not a method call, an \emph{if statement} cannot be considered as a sink when using state-of-the-art static taint analyzers~\cite{10.1007/978-3-642-30921-2_17, arzt2014flowdroid, 10.1145/2660267.2660357}.
A concrete example of what \difuzer tracks is given in Listing~\ref{code:instrumentation}.
On line 7, \emph{countryCode} variable is tainted from \emph{getNetworkCountryIso()} source.
This value is then used (line 9) to perform a test and trigger malicious activity (line 9).
As an \emph{if statement} is not considered a sink, a flow cannot be found.

\begin{listing}[h]
\begin{minted}{java}
public void method() {
  String b = Build.BRAND;
+ b = BuildClass.getBRAND(); // dummy method call for field access
  String p = Context.TELEPHONY_SERVICE;
  Object o = this.getSystemService(p);
  TelephonyManager tm = (TelephonyManager) o;
  String countryCode = tm.getNetworkCountryIso();
+ IfClass.ifMethod(countryCode, "RU"); // dummy method call for if statement
  if(countryCode.equals("RU")){ performMaliciousActivity(); }
}
\end{minted}
    \caption{Example of app instrumentation performed by \difuzer (Lines with
”+” represent added lines).}
    \label{code:instrumentation}
\end{listing}

Our approach overcomes this limitation by instrumenting apps.
To accomplish this, the app code is first transformed into Jimple~\cite{vallee1998jimple}, the internal representation of Soot~\cite{vallee2010soot}.
Then, \difuzer iterates over every condition of the app, and for each condition, \difuzer inserts a dummy method \texttt{ifMethod} with the variables involved in the condition as parameters.
This \texttt{ifMethod()} is static and declared in a dummy class \texttt{IfClass} that contains all instrumented methods related to conditions.
See line 8 in Listing~\ref{code:instrumentation}.

Once the instrumentation is over, we dynamically register every newly generated method calls as sinks to \flowdroid.

\noindent
\textbf{Sources related challenge:} 
As described in Section~\ref{sec:sources_sinks}, we consider, in this study, \texttt{Build} class' fields as sources.
Since field accesses are not method calls, we follow the same process as for \emph{if statements} to insert dummy methods. 
More specifically, \difuzer generates a static method call on-the-fly representing a field access from the \texttt{Build} class.
Listing~\ref{code:instrumentation} depicts an example of this instrumentation process, where the dummy method \texttt{getBRAND()} of the dummy class \texttt{BuildClass} is inserted in line 3.
Furthermore, newly generated method calls are registered as sources for taint tracking.

\subsection{Anomaly detection}
\label{sec:anomaly_detection}

This section presents \difuzer's second module, which relies on anomaly detection.
In particular, we detail the unsupervised machine learning technique used to detect abnormal triggers. 

\vspace{-.1cm}

\subsubsection{Why a One-Class SVM?}
\label{sec:why_oc_svm}

A classical classification problem requires samples from positive and negative classes to build a model, which is then used to assign labels to test instances~\cite{kotsiantis2007supervised}.
This induces possessing a reasonable amount of samples from two classes, which is not the case in our study. Indeed, the SHSO detection problem is challenging, and to the best of our knowledge, there is no ground truth made publicly available.
Thus, using supervised learning in our study is not practical and present limited feasibility.

Therefore, we decided to rely on an unsupervised learning technique to detect SHSOs, particularly on a One-Class Support Vector Machine (OC-SVM) machine learning technique. 
An SVM algorithm was chosen due to its ability to generalize~\cite{788640} and its resistance to over-fitting~\cite{xu2009robustness}.
The general idea of OC-SVM is to identify the smallest hyper-sphere to include most of the samples of the positive samples~\cite{958946}.
A sample considered as an outlier by the model means the data-point is not in the hyper-sphere.

\subsubsection{Features extraction}
\label{sec:features_extraction}

As already said, the second \difuzer module's objective is to detect abnormal triggers with the intuition that these triggers are HSOs for which the likelihood to be a logic bomb is high, namely SHSOs.
This module implements an OC-SVM algorithm which takes as input feature vectors computed from the triggers previously extracted from the entry-points yielded by the first module of \difuzer (cf. Figure~\ref{fig:overview}).

To engineer anomaly detection features, we reviewed surveys~\cite{zhou2012dissecting,7828100} and related-papers~\cite{ALAM2017230,10.1007/978-3-319-24018-3_12,6906344,pan2017dark} discussing Android malware and investigated the techniques used by malware writers to hide malicious code within apps. 
Eventually, we identified nine unique trigger/behavior features that are described in the following.

In the remainder of this section, we consider a trigger $\tau = (c, T_c, \Phi_c)$ and its guarded code $\Gamma = T_c \cup \Phi_c$ (cf. Section~\ref{sec:background_definition}).

\difuzer builds a feature vector $v = <S, N, D, R , B, P, M_1, S_1, J>$ for a given trigger where:

\noindent
\textbf{S: Number of sensitive methods used in guarded code. } 
Intuitively, this feature represents how much a trigger controls the execution of sensitive methods. 
Indeed, while HSOs guard the execution of sensitive operations for performing sensitive activities~\cite{7546513}, benign triggers, in the general case, perform benign activities, i.e., invoke few sensitive methods, or not at all.
To retrieve this value, \difuzer iterates over every statement of $\Gamma$ and recursively checks whether a sensitive method is called or not.
For this purpose, we gathered a list of sensitive APIs constructed in previous work~\cite{10.1145/2382196.2382222}.

\noindent
\textbf{N: Is native code used in guarded code? } 
Since analyzing native code is more challenging than Java bytecode~\cite{LI201767}, Android malware developers tend to hide malicious code from automated analyses in native code~\cite{ALAM2017230,10.1007/978-3-319-24018-3_12}.
Hence, this feature is a boolean value that, when set to 1, means native code is used in $\Gamma$, 0 otherwise.

\noindent
\textbf{D: Is dynamic loading used in guarded code? } 
Dynamic class loading is not exclusively used in malware. 
However, as malware is becoming increasingly sophisticated, they use built-in capabilities like dynamic loading to hide from automated analyses~\cite{6906344}.
Consequently, likewise native code, this feature is a boolean value set to 1 if dynamic loading is used in $\Gamma$, 0 otherwise.

\noindent
\textbf{R: Is reflection used in guarded code? } 
Android malware writers tend to use more and more reflection-based code~\cite{6906344} since most of the state-of-the-art techniques overlook this property due to the challenging task of resolving it.
Therefore, this feature is set to 1 if reflection is used in $\Gamma$, 0 otherwise.

\noindent
\textbf{B: Does guarded code trigger background tasks? } 
Android apps rely on the Service component to run background tasks.
Hence, with this feature, we aim at capturing the fact that the app under analysis performs stealthy operations without user knowledge. 
The intuition here is that SHSOs' role is to hide code both from security analysts and end-users (e.g., in the case of a logic bomb).
This feature is set to 1 if background services are triggered in $\Gamma$, 0 otherwise.

\noindent
\textbf{P: Are parameters of condition used in guarded code?} 
This feature captures the dependency of a condition to its guarded code.
The hypothesis is that, in the case of SHSOs, the guarded code does not use values used in the condition since they represent different behaviors.
To achieve this, \difuzer performs a def-use analysis of the guarded code to verify if any variable used in the condition is used before being assigned a new value.
If this is the case, the feature is set to 1, 0 otherwise.

\noindent
\textbf{M$_1$: Number of app methods called only in guarded code. } 
With this attribute, we attempt to uncover the number of methods defined in the app called only in the guarded code of a trigger.
The rationale is that app methods that are only used under a specific circumstance are likely to be defined only for this specific circumstance, representing hidden behavior~\cite{fratantonio2016triggerscope}.
To retrieve this number, \difuzer queries the call-graph (built using SPARK~\cite{10.1007/3-540-36579-6_12} algorithm) for each method call in the guarded code to verify if it has only one incoming edge (i.e., it is only called within the current method).

\noindent
\textbf{S$_1$: Number of sensitive methods called only in guarded code. } 
In the same way as M$_1$, we aim to capture the number of sensitive methods only used in the guarded code of a given trigger.

\noindent
\textbf{J: Behavior difference between branches.} 
Intuitively, two bran\-ches of an SHSO should be noticeably different. Indeed, of the two branches, one is considered the normal behavior (no or few sensitive operations) if the condition is not satisfied and the other as the sensitive behavior (sensitive operations) if the condition is satisfied~\cite{pan2017dark}.
Therefore, to compute this difference, \difuzer first inter-procedurally retrieves sensitive method calls in both branches of a given trigger.
Let $X_{T_c}$ and $X_{\Phi_c}$ respectively be the sets of sensitive methods in the true and the false branch of a trigger.
Therefore, to compute this difference of the two branches, \difuzer relies on the Jaccard distance: $D_j(X_{T_c}, X_{\Phi_c}) = 1 - \frac{\lvert X_{T_c} \cap X_{\Phi_c}\rvert}{\lvert X_{T_c} \cup X_{\Phi_c}\rvert}$, which characterizes the behavior difference of the two branches.
A value close to 1 means that both branches are dissimilar.

\subsubsection{Training phase}
\label{sec:learning}
To train our OC-SVM model, we need samples of a positive set, i.e., triggers considered normal.
Therefore, we randomly chose \num{10000} goodware (i.e., VirusToal~\cite{total2012virustotal} score = 0) from \az~\cite{Allix:2016:ACM:2901739.2903508}.
Then, for each of these apps, we applied \difuzer to extract a feature vector for each app's condition.

Afterward, we randomly chose \num{10000} feature vectors\footnote{The number of extracted vectors is orders of magnitude higher.
However, for efficiency, we validated that a random set of \num{10000} vectors yields an acceptable performance.} from those yielded by \difuzer, which we labeled as positive (i.e., part of the normal behavior).
We then trained our One-Class Classification based anomaly detector, leveraging LibSVM~\cite{10.1145/1961189.1961199}.
To ensure that the selected training set does not bias the trained model's performance, we split it and compute Accuracy in 10-fold cross-validation. Overall, we achieve a stable Accuracy of \num{99.91}\% on average. 

\section{Evaluation}
\label{sec:evaluation}

To evaluate \difuzer, we address the following research questions:

\noindent
\textbf{RQ1:} What is the performance of \difuzer for detecting Suspicious Hidden Sensitive Operations (SHSOs) in Android apps?

\noindent
 \textbf{RQ2:} Can \difuzer be used to detect logic bombs? We address this question by considering three sub-questions:
    \begin{itemize}
    \item  \textbf{RQ2.a:}  Are SHSOs detected by \difuzer likely logic bombs?
    \item  \textbf{RQ2.b:} How does \difuzer compare against \ts, a state of the art logic bomb detector?
    \item  \textbf{RQ2.c:} From a qualitative point of view, does \difuzer lead to the detection of non-trivial triggers/logic bombs?
\end{itemize}
\noindent
 \textbf{RQ3:} Can SHSO detection in goodware reveal suspicious behavior?

\subsection{RQ1: Suspicious Hidden Sensitive Operations in the wild}
\label{sec:wild}

In this section, we assess the efficiency of \difuzer to find SHSOs on a dataset of malicious applications.

\noindent
\textbf{Dataset.}
To the best of our knowledge, there is no SHSO ground-truth available in the literature. 
Consequently, in this study, we considered \num{10000} malicious Android apps as our malicious dataset. 
These apps were released in 2020, collected from the \az~\cite{Allix:2016:ACM:2901739.2903508} repository, and have been flagged as malware by at least five antivirus scanners in VirusTotal. 

We contacted the authors of state of the art approaches (e.g.,  \darkhazard~\cite{pan2017dark}, and \ts~\cite{fratantonio2016triggerscope}) to get their artifacts (data\-sets and tools) for comparative assessment. 
Unfortunately, no artifact was made available to us.

\noindent
\textbf{Libraries.}
It has been shown in the literature~\cite{li2016-libraries,7546512} that library code can affect analyses performed over Android apps since it often accounts for a larger part than the app's core code.
Consequently, in this study, we considered two cases: 
(1) with-lib analysis (i.e., we consider the entire app code including library code);
(2) without-lib analysis (i.e., we consider only developer code).
To rule out libraries, we rely on the state-of-the-art list available in ~\cite{li2016-libraries}.

\noindent
\textbf{Post-Filter.}
As a precaution, before analyzing the results without libs, we listed the classes in which \difuzer found potential sensitive triggers to search for redundant classes that could indicate libraries. 
We were able to filter out \numberLibFiltered additional libraries that were not listed in the list we used and provided by~\cite{li2016-libraries}.

In the following, when referring to the analysis without libraries, we consider the \numberLibFiltered libraries previously presented as well as the libraries of the list in~\cite{li2016-libraries} as filtered. It accounts for a total of \totalLibs library classes and packages filtered.

\subsubsection{Efficiency of Detecting SHSOs}
\label{sec:effiency}
We recall that \difuzer is targeted at detecting SHSOs. While in RQ2 we investigate the likelihood for these SHSOs to be logic bombs, we first investigate the efficiency (with RQ1) of \difuzer in the detection of SHSOs. We further perform an ablation study to highlight the performance of the anomaly detection module.

\begin{figure}[h]
    \centering
    \includegraphics[width=.7\linewidth]{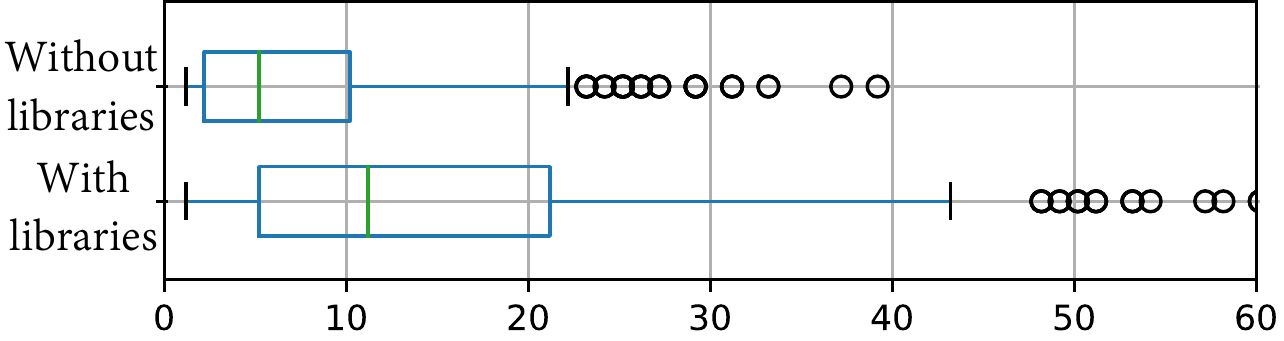}
    \caption{Distribution of the number of SHSO(s) per app in analyses with and without libraries (only apps with at least one SHSO are considered).}
    \label{fig:distribution_triggers_holistic_partial}
\end{figure}

In Table~\ref{table:results}, we report the results of applying \difuzer (with the anomaly detection step activated) on our \num{10000} malware dataset. 
When analyzing the entire apps, \difuzer detects at least one SHSO in \appsWithTriggersHolistic apps (\appsWithTriggersHolisticPercent\%).
Overall, \difuzer detects \numberTriggersHolistic SHSOs in these 339 apps leading to an average number of \numberTriggerPerAppHolistic SHSOs per app.  
In comparison, when only the app developers' code is considered, \difuzer detects at least one SHSO in \appsWithTriggersPartial apps (\appsWithTriggersPartialPercent\%), with a total number of \numberTriggersPartial SHSOs detected and an average number of \numberTriggerPerAppPartial SHSOs per app.
We note that the \num{3437} (\numberTriggersHolistic-\numberTriggersPartial) SHSOs that are not in the app developer code, are actually detected in 68 libraries suggesting that only a few libraries contain SHSOs .
Figure \ref{fig:distribution_triggers_holistic_partial} further details the distribution of detected SHSOs per apps. 

\emph{These first results show that SHSOs indeed exist in malicious apps, but in relatively low number (in around 3\% of the apps). 
However, when SHSOs are present in an app, they are not rare (on average, about 8 SHSOs per app in the developer code). 
Finally, SHSOs are more prevalent in library code than in app developer code, but only a few libraries contain SHSOs.}

Table~\ref{table:results} also reports the average numbers of triggers before and after applying the anomaly detection step (i.e., the second module of \difuzer). 
Interestingly, we can see that this anomaly detection drastically reduces the number of triggers that are considered as SHSOs.
Indeed, when considering the \num{10000} apps, there are on average $ 174336/10000 \approx$ \flowBeforeADHolistic and $ 146018/10000 \approx$ \flowBeforeADPartial triggers per apps (with or without libraries respectively) generated by the first module of \difuzer, i.e., by the taint analysis step.  
After the anomaly detection step, these numbers drop to $ \numberTriggersHolistic/10000 \approx$ \flowAfterADHolistic and  $ \numberTriggersPartial/10000 \approx$ \flowAfterADPartial respectively, corresponding to a decrease of 96\% and 98\% respectively.

\emph{These results show that the anomaly detection step has a significant impact on the number of detected SHSOs by significantly reducing the search space of triggers by up to 98\%. This search space reduction is key when the ultimate goal is to detect malicious code and to support security analysts manual inspection (cf. Section~\ref{sec:logic_bomb_detection}).}

\begin{table}[ht]
    \centering
    \caption{Results of the experiments executed on \num{10000} malware with and without taking into account libraries.}
    \begin{adjustbox}{width=\columnwidth,center}
    \begin{tabular}{l|p{23mm}|p{27mm}}
        & \centering Analysis with libs & Analysis without libs \\ 
        \hline
        Number of apps with SHSO(s) & \appsWithTriggersHolistic & \appsWithTriggersPartial \\
        Number of SHSOs & \numberTriggersHolistic & \numberTriggersPartial \\
        Number of SHSOs/app & \numberTriggerPerAppHolistic & \numberTriggerPerAppPartial \\
        Average \# triggers (i.e., before Anomaly detection) & \flowBeforeADHolistic & \flowBeforeADPartial \\
        Average \# SHSOs (i.e., after Anomaly detection) & \flowAfterADHolistic & \flowAfterADPartial \\
        Mean analysis time & \analysisTimeHolistic s & \analysisTimePartial s\\
        \hline
    \end{tabular}
    \end{adjustbox}
    \label{table:results}
\end{table}

We further inspect the SHSOs detected by \difuzer by focusing on the app developer code only (we do not consider library code). 
Table~\ref{table:type_triggers} lists the top 10 types of trigger that \difuzer was able to discover. 
The second column gives some examples of methods considered sources for the taint tracking to uncover SHSO entry-points.
We note the diversity of types of triggers that developers use. 
For instance, a developer can decide to trigger (or not) the sensitive code if: 
(Database trigger type) specific values are present in databases (e.g., contacts, messages); 
(Internet trigger type) external orders say so; 
(Build, Telephony, and Camera trigger types) the device is not an emulator; 
(Connectivity, and Wi-Fi trigger types) the device has Internet access; 
(Location rigger type) the user is in a pre-defined location; 
Note that the methods in Row 3 have been dynamically generated by \difuzer during instrumentation to track the Build class's field values.

\begin{table}[ht]
    \centering
    \caption{Top ten trigger types discovered by \difuzer in the developer code. \normalfont{(T. = Triggers)}}
    \begin{adjustbox}{width=\columnwidth,center}
    \begin{tabular}{l|l|l||l|l|l}
        Trigger Type & Examples of methods & \# T. & Trigger Type & Examples of methods & \# T.\\ 
        \hline
        Database & getString, getInt, getCount & 785 & Location & getLastKnownLocation, getLongitude & 84 \\
        Internet & getResponseCode, getResponseMessage & 715 & Wi-Fi & isWifiEnabled, getConnectionInfo & 76 \\
        Build & getMODEL, getMANUFACTURER & 374 & Power & isScreenOn, isInteractive & 47 \\
        Telephony & getDeviceId, getNetworkOperatorName & 97 & Audio & getStreamVolume, isMusicActive & 37 \\
        Connectivity & getActiveNetworkInfo, getNetworkInfo & 88 & Camera & getCameraIdList & 28 \\
        \hline
    \end{tabular}
    \end{adjustbox}
    \label{table:type_triggers}
\end{table}

Regarding the component types in which \difuzer found SHSOs, 90\% of SHSOs are in methods of "normal" classes, i.e., not Android components.
SHSOs are found in \texttt{Activities} in 9\% of the cases.
However, they are rarely found in \texttt{Services} and \texttt{Broadcast Receivers} (less than 1\%).

\subsubsection{Manual Analyses}
\label{sec:manual}
Since static analysis approaches often suffer from false alarm issues, i.e., they report a large proportion of false-positive results, 
we decided to verify the detection capabilities of \difuzer manually. 
To that end, the authors of this paper randomly selected a statistically significant sample of \numberAppsManuallyAnalyzed apps out of the \appsWithTriggersPartial apps in which SHSOs exist in developer code, with a confidence level of 99\% and a confidence interval of $\pm$ 10\%. 
Only one sample was found to be a false-positive result.
Indeed this app verifies if it is running in an emulator by comparing \texttt{Build.PRODUCT}, \texttt{Build.MODEL}, \texttt{Build.MANUFACTURER}, and \texttt{Build.HARDWARE} against well-known strings such as "generic", "Emulator", "google\_sdk", etc.
This test seems sensitive, but the guarded code displays the following message to the user: "Scooper Warning: App is running on emulator.". 
Therefore, \difuzer achieves a precision of \difuzerPrecision\% to find \emph{Suspicious Hidden Sensitive Operations} on this dataset.
We release the annotated list of \numberAppsManuallyAnalyzed apps that were manually checked for transparency in the project's repository.

\subsubsection{Analysis Time}
\label{sec:analysisTime}
The last row in Table~\ref{table:results} reports \difuzer analysis time. 
\difuzer outperforms state-of-the-art trigger detectors with an average of \analysisTimePartial s per app (\analysisTimeHolistic s for the analysis with libraries, with an average DEX size of \num{7.03} MB per app), making \difuzer suitable for large-scale analyses. 
In comparison, state-of-the-art tools such as \ts~\cite{7546513} and  \darkhazard~\cite{pan2017dark}) require \num{219.21} s and \num{765.3} s per app respectively. 
Note that \num{85.42}\% (i.e., \num{28.65} seconds on average) of this time is reserved for the taint analysis. 
Also, \numberAppsTimeoutPartial apps (\numberAppsTimeoutPartialPercent\%) reached the timeout (i.e., 1 hour) before the end of the analysis.

\highlight{
\textbf{RQ1 answer: }\difuzer detects SHSOs in Android malware with high precision, i.e., \difuzerPrecision\% in less than 35 seconds on average. Among the average 14.6 HSOs identified in an app based on triggers spotted by static taint analysis, only 2\% are suspicious according to anomaly detection, which shows that \difuzer is effective in reducing the search space for manual analysis.
}

\vspace{-.2cm}

\subsection{RQ2: Can  \difuzer detect  logic  bombs?}
\label{sec:logic_bomb_detection}

In this section, we \ding{182} evaluate \difuzer's efficiency in detecting logic bombs (RQ2.a), \ding{183} compare it against \ts (RQ2.b), and \ding{184} discuss logic bomb use cases in real-world apps (RQ2.c).

\subsubsection{RQ2.a: Are SHSOs detected likely to be logic bombs?\label{sec:eval:logic_bombs}} \textcolor{white}{x}

Until now, we have shown that \difuzer is effective in detecting SHSOs.
From a security perspective, however, we must further show that these SHSOs are actually malicious. In other words, are these SHSOs likely to be logic bombs. Unfortunately, such assessment is challenged by the lack of ground truth in the literature. We therefore require extra manual analysis effort of reported results.

\textbf{Initial Manual Analysis:}
In previous Section~\ref{sec:manual}, we present our manual analysis of SHSOs detected in \numberAppsManuallyAnalyzed apps.  
During this analysis, we further checked if the detected SHSOs contain malicious code. 
In particular, for each app under analysis, we gathered information about the reason it was flagged by antiviruses (e.g., on VirusTotal).
Then, in the guarded code of the potential SHSO found by \difuzer, we looked for malicious behavior matching our information previously gathered.
For instance, if: (1) an app is labeled as being a trojan stealing the device's information; (2) the potential SHSO is performing emulator detection (e.g., calling \texttt{System.exit()} method if the device is running in an emulator); and (3) the behavior exhibited in the code guarded by the condition detected by \difuzer is gathering the device's information (e.g., unique identifier, current location, etc.) and sending it outside the device, the SHSO is considered a logic bomb.

Eventually, \numberLogicBombsInManual apps (i.e., \numberLogicBombsInManualPercent\%) were manually confirmed to be logic bombs, i.e., the SHSOs were triggering malicious code. 

\textbf{Semi-Automated further Analysis: }
Manual investigation is time-consuming. 
This is the reason why we inspected \numberAppsManuallyAnalyzed apps and not all \appsWithTriggersPartial apps reported to having at least one SHSOs within the developer code parts. 
To quickly enlarge the set of identified logic bombs, we decided to follow a simple but efficient process.  
It is known that malicious developers often reuse the same piece of code in different apps~\cite{7828100}.
Therefore, for each already identified logic bomb, we search for similarities (i.e., SHSOs found in the same class name, same method name, and the same type of trigger used) in SHSOs contained in the 157 ($\appsWithTriggersPartial - \numberAppsManuallyAnalyzed$) remaining apps. 
Our analysis yielded \numberAdditionalLogicBombs additional apps containing logic bombs that were manually verified and confirmed.
Eventually, our logic bomb dataset, called \db, contains \numberLogicBombsInDataset Android apps, each with an identified logic bomb.
We believe this dataset to be useful to the community to further improve logic bomb detection in Android apps.
We made it publicly available in the project's repository.

\textbf{Discussion about HSO, SHSO and Logic Bomb:} In the literature 
~\cite{pan2017dark,fratantonio2016triggerscope}, HSO is consistently defined as a sensitive operation that is hidden by specific triggering conditions. Nevertheless, the notion of ``sensitive operation'' is not clearly delineated, which challenges comparison across approaches.
In our work, we postulate that while detecting HSOs is an important first step, it is not enough to help security analysts. 
Indeed, as shown by our manual analysis, a large proportion of HSOs are indeed sensitive but not necessarily suspicious. 
As a result, most of the detected HSOs are legitimate and do not require any inspection effort from security analysts.

In this context, if the goal is to detect real security issues and reduce the burden of security analysts, a tool such as  \darkhazard~\cite{pan2017dark} which detects {\em HSOs} in \num{18.7}\% of apps within a set of over \num{300000} apps (including malicious and benign apps) appears to be unpractical. 
In contrast, \difuzer detects {\em suspicious HSOs} in \appsWithTriggersHolisticPercent\% of the analyzed apps (when libraries are considered), and our manual analyses confirm that in about 30\% of the apps, these SHSOs are logic bombs, making the work of security analysts easier.
Though both \darkhazard dataset and our dataset are different (we were not able to get the \darkhazard's authors dataset), if we compare the \num{18.7}\% of apps with HSOs reported by \darkhazard, with the \appsWithTriggersHolisticPercent\% reported by \difuzer, we can say that \difuzer  
reduces the search space by up to \searchSpaceReduced\% ($(\num{18.7} - \num{3.39}) \times \frac{\num{100}}{\num{18.7}} = \searchSpaceReduced$) to accelerate the identification of logic bombs.

\highlight{
\textbf{RQ2.a answer: }
By triaging HSOs to focus on suspicious ones based on anomaly detection, \difuzer was able to reveal 30 logic bomb instances in a sampled subset of malware apps having SHSOs.
Besides, we release \db, an annotated dataset of \numberLogicBombsInDataset Android apps confirmed to be using logic bombs.
}

\subsubsection{RQ2.b: How  does \difuzer compare  against \ts, a state of the art logic bomb detector?}
\label{sec:eval:triggerscope}
\textcolor{white}{x}

In the absence of a public ground-truth for Android logic bomb instances,  we perform experimental comparisons against the \ts state-of-the-art detector in the literature that relies on static analysis. 
Although \ts is not publicly available, we are able to build on a replication based on technical details provided in \ts paper~\cite{fratantonio2016triggerscope}.

Overall, our approach differs from \ts's by three major differences: \ding{182} \textbf{Technique}: \ts uses symbolic execution to tag variables with a limited number of values, we use static data flow analysis; \ding{183} \textbf{Target}: \ts detects hidden sensitive operations (i.e., whether at least one sensitive method is called within the guarded code of a trigger), whereas \difuzer’s goal is to detect suspicious hidden sensitive operations (i.e., the guarded code is sensitive and implements an abnormal behavior); and \ding{184} \textbf{Approach}: \ts maintains a list of sensitive methods and uses the occurrence of any of them as the sole criterion, \difuzer implements an anomaly detection scheme where the presence of sensitive methods is one feature among many others. While \ts and \difuzer both rely on list of sources to find triggers of interest, \ts handpicks a limited set of methods, whereas \difuzer’s list is based on a systematic mapping (cf. Section~\ref{sec:sources_sinks} - we leverage patterns to systematically search for sources).

\noindent
\textbf{\em Does \ts identify as logic bombs the SHSOs flagged by \difuzer?}

\begin{figure}[h]
    \centering
    \includegraphics[width=.49\linewidth]{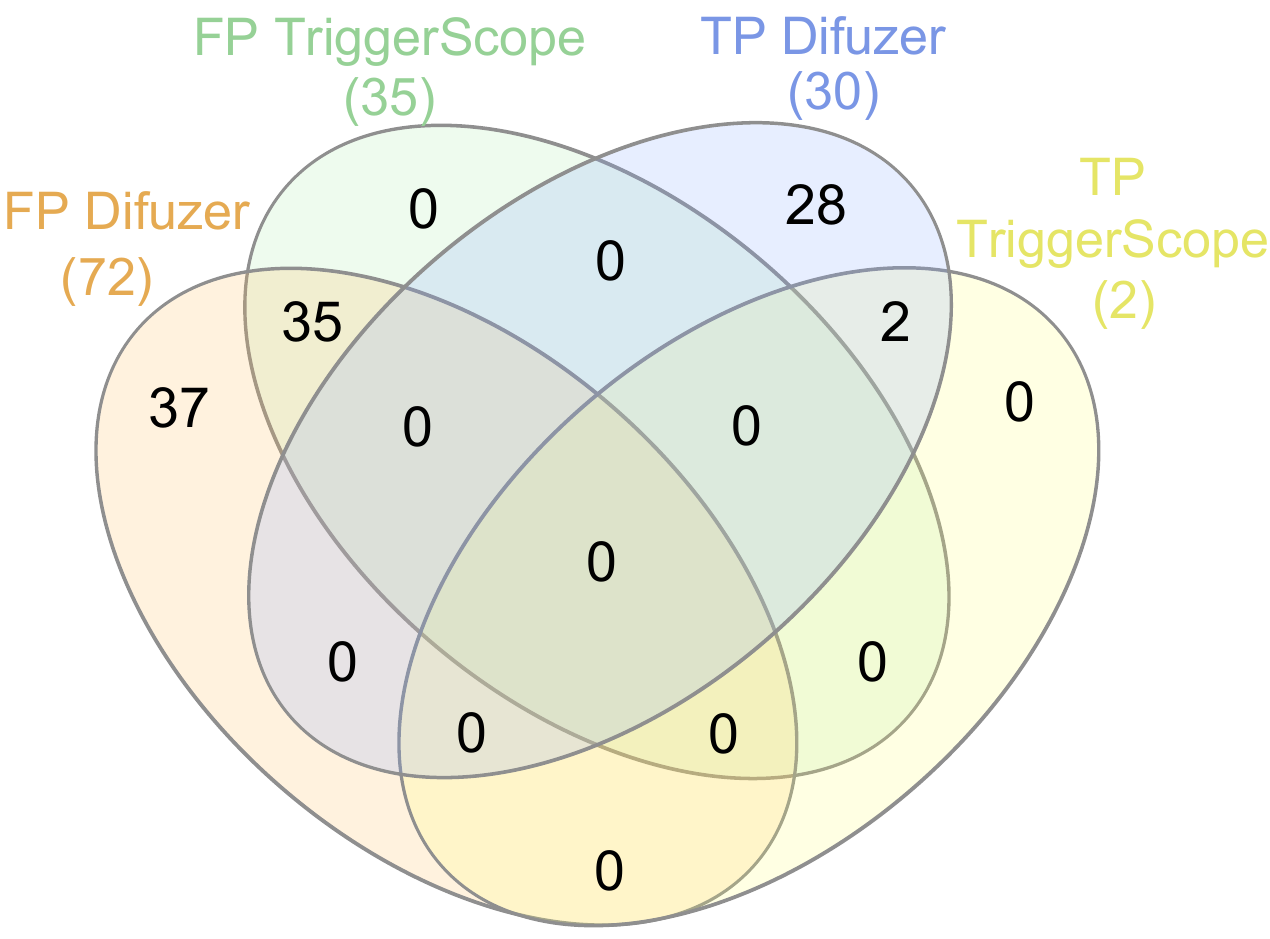}
    \includegraphics[width=.49\linewidth]{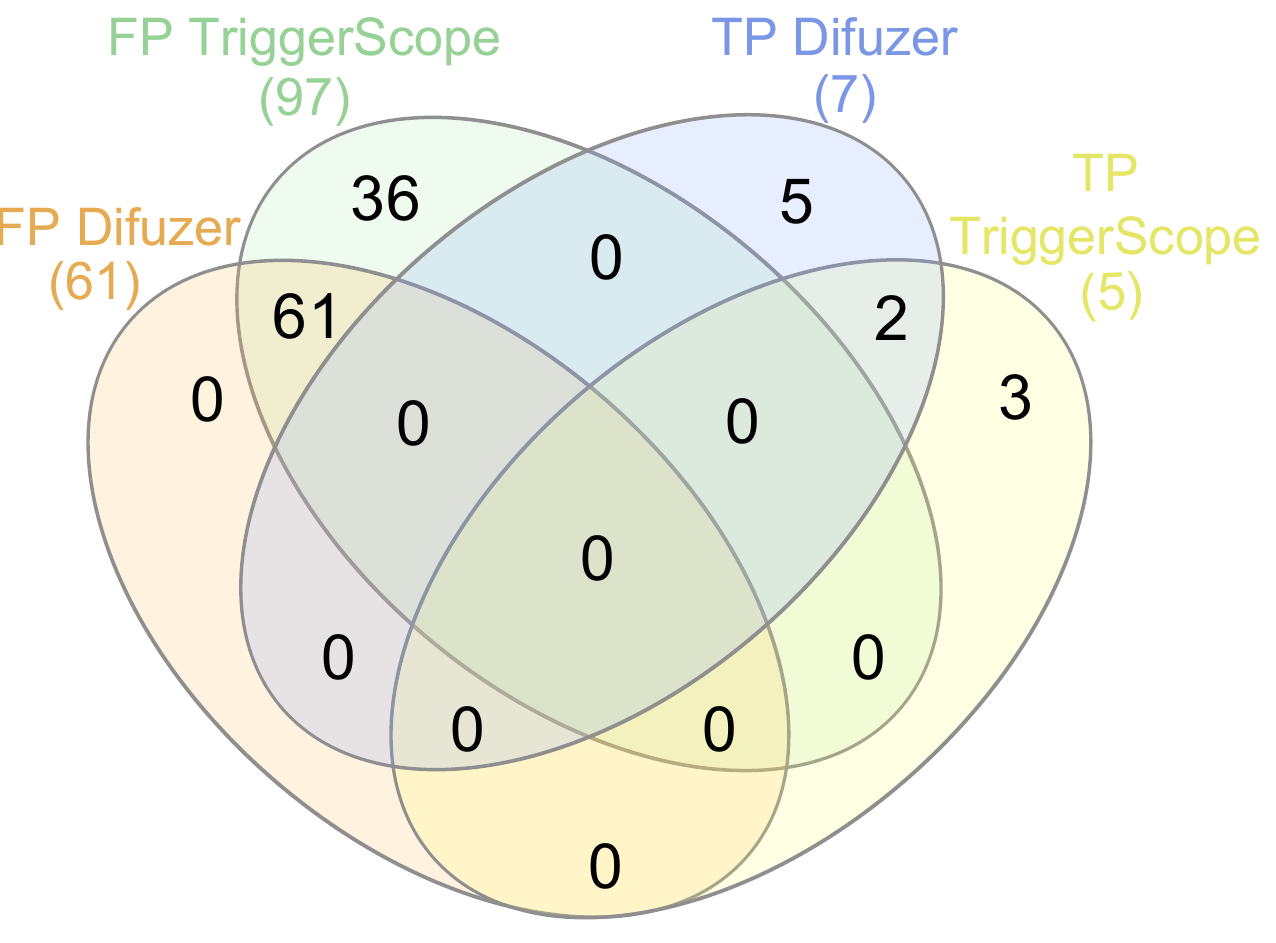}
    \caption{Venn Diagram representing results of \ts and \difuzer on 102 apps originally detected by \difuzer on the left, and \ts on the right. \normalfont{(FP = False Positive, TP = True Positive)}}
    \label{fig:venn2}
\end{figure}

We applied \ts on the subset of \numberAppsManuallyAnalyzed apps where \difuzer identified a SHSO (cf. Section~\ref{sec:eval:logic_bombs}). 
The objective is to check whether \ts is more or less accurate than \difuzer. Typically, among the 30 logic bombs that have been manually verified as true positives, how many are detected by \ts. Similarly, does \ts detect logic bombs (manually verified as true positives) that \difuzer could not. Figure~\ref{fig:venn2} illustrates the differences in logic bomb detection (left figure). Overall:

\begin{itemize}[leftmargin=*]
    \item \ts did not flag any logic bomb that \difuzer did not.
    \item \ts could only detect 2 logic bombs among the 30 logic bombs that \difuzer correctly identified.
    \item As reported in the literature~\cite{samhi2021ineffectiveness}, \ts exhibits a very high false positive rate at 94.6\%: 35 among its 37 detections are false positives (the rate for \difuzer is 70.6\%, 72/102).
\end{itemize}

\noindent
\textbf{\em Does \difuzer fail to flag as SHSOs the logic bombs detected by \ts?}

We recall that, contrary to \difuzer, which builds on anomaly detection, \ts is restricted to detect only logic bombs where the trigger involves location-, time-, and SMS-related properties.
Aligning with the assessment of \difuzer, we applied \ts on our set of \num{10000} malware. 
\ts reported 591 logic bombs in 149 apps ($\sim$4/app):  \num{98.6}\% of the reported cases are time-related. In the absence of ground truth, we again propose to manually verify a random sample set of reported logic bombs.
To facilitate comparison with \difuzer, we sample \numberAppsManuallyAnalyzed apps (we simply considered the same number of apps as in the previous question), and manually confirmed that for 97 (\num{95.1}\%) apps, the reported logic bombs are false positives. In 5 (\num{4.9}\%) apps, we found at least one reported logic bomb to be a true positive. 

We further check whether on these 102 apps where \ts reported a logic bomb, \difuzer also flags any case of SHSO:
\difuzer flagged 68 apps as containing SHSOs, among which 7 are manually confirmed to be logic bombs.
The details of the comparison between \ts and \difuzer are presented in the Venn Diagram in Figure~\ref{fig:venn2} (right figure). 
We note that: 

\begin{itemize}[leftmargin=*]
    \item 2 logic bombs are detected by both \difuzer and \ts.
    \item 5 SHSOs detected by \difuzer are actual logic bombs, but not detected by \ts. Indeed, \ts is limited by its focus on time, location and SMS-related triggers.
    \item 3 logic bombs are detected by \ts, but not detected by \difuzer.  Our prototype implementation considers a limited list of sources, which do not cover those 3 logic bomb cases.
\end{itemize}

Although we do not have a complete ground truth (with information about all cases of logic bombs), confirming and comparing detection reports by \difuzer and \ts offers an alternative to assess to what extent each may be missing some logic bombs. The results described above suggest that \difuzer suffers significantly less from false-negative results than \ts.

\highlight{
\textbf{RQ2.b answer: }
Overall, \difuzer outperforms \ts by
detecting more logic bombs more accurately (wrt. false positives), and by
missing less logic bombs (wrt. false negatives).
}

\subsubsection{RQ2.c: From a qualitative point of view, does \difuzer lead to the detection of non-trivial triggers/logic bombs?}
\label{sec:case_studies}
\textcolor{white}{x}

In this section, we discuss two real-world apps in which \difuzer revealed logic bombs that cannot be detected by \ts.

\noindent
\textbf{Advertisement Triggering.} \difuzer revealed an interesting logic bomb in "com.walkthrough.knife.assassin.hunter.baoer" app which is an adware app of the HiddenAd family.
The app uses the \texttt{android.\-app.job.JobService} class of the Android framework to schedule the execution of jobs (the developer can handle the code of the job in \texttt{onStartJob} method).
In the \texttt{onStartJob} method, the app takes advantage of the \texttt{PowerManager} of the Android framework to check if the device is in an interactive state (i.e., the user is probably using the device) with method \texttt{isScreenOn()}.
If this is the case, the app displays advertisements to the user and schedules the same class's execution after a certain time.

\noindent
\textbf{Data Stealer.} Logic bombs can also be used to trigger data theft under the condition that the data is available.
For instance, in app "com.magic.clmanager", which is a Trojan (hidden behind a cleaning app) capable of stealing data on the device, \difuzer found a logic bomb related to the device unique identifier.
Indeed, in method \texttt{d(Context c)} of the class \texttt{c.gdf}, a check is performed against the value returned by method \texttt{getDeviceId()} to verify if the value matches specific values (emulator detection) in a given file named "invalid-imei.idx".
In the case the app considers that the device is not an emulator, it triggers the stealing of sensitive information about the device such as the current location, phone number, information on the camera, information about the Bluetooth, disk space left, whether the device is rooted or not, the current country, the brand, the model, information about the Wi-Fi, etc.
Afterward, this information is written in a file and sent to a native method for further processing.

\subsection{RQ3: SHSOs in benign apps}
\label{sec:goodware}

Until now, we have focused on malware.
However, SHSOs are not exclusively found in malicious apps~\cite{pan2017dark}.
Therefore, in this section, we intend to conduct a study on benign applications.

\textbf{Results.} As confirmed in Section~\ref{sec:effiency} and in previous studies~\cite{pan2017dark,fratantonio2016triggerscope}, benign libraries and benign Android apps implement HSOs. Our study confirms this finding. 
Even more, \benignWithTriggersPartial benign apps (\benignWithTriggersPartialPercent\%) were flagged by \difuzer to contain suspicious HSOs.
We further manually analyzed 20 apps randomly selected from our results and confirmed that they all contain at least one SHSO.
Table~\ref{table:top_ten_benign} shows the different trigger types used in benign apps to trigger SHSOs.
A significant result here is that benign apps use considerably less the "Build" trigger type (see Table~\ref{table:type_triggers} for comparison) than malicious apps.
Similarly, the "Telephony" trigger type is less used in benign apps than in malicious apps.
This induces that, in benign apps, decisions are less taken depending on values derived from methods like: \texttt{getDeviceId()}, \texttt{getNetworkOperatorName()}, \texttt{getPhoneType()}, \texttt{getMODEL()}, \texttt{getMANUFACTURER()}, or \texttt{get-}\linebreak \texttt{FINGERPRINT()}.
A hypothesis would be that benign apps are less prone to recognize an emulator environment (and use this information to set triggering conditions).

\begin{table}[h]
    \centering
    \caption{Top ten trigger types used by benign Android apps.}
    \begin{adjustbox}{width=\columnwidth,center}
    \begin{tabular}{c|c|c|c|c|c|c|c|c|c}
        Database & Internet & Location & Connectivity & Audio & Telephony & Wi-Fi & View & Activity & Build \\
        \hline
        897 & 283 & 264 & 74 & 63 & 58 & 25 & 21 & 19 & 19 \\
    \end{tabular}
    \end{adjustbox}
    \label{table:top_ten_benign}
\end{table}

\begin{figure}[h]
    \centering
    \includegraphics[width=.8\linewidth]{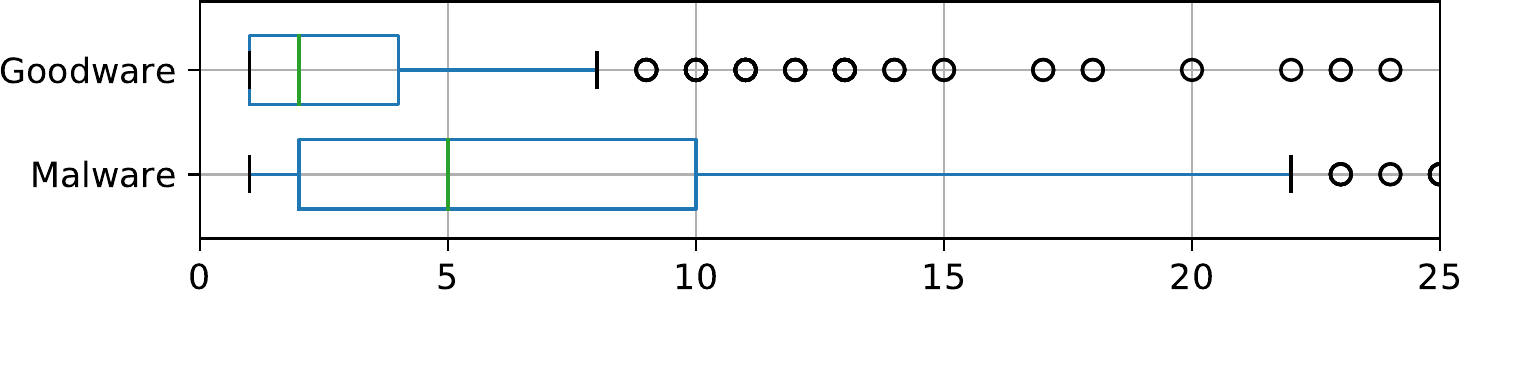}
    \caption{Distribution of the number of SHSO(s)/app in goodware and malware \normalfont{(apps with at least 1 SHSO are considered).}}
    \label{fig:distribution_triggers}
\end{figure}

Besides, we can see in Figure~\ref{fig:distribution_triggers} that, in comparison with malicious apps, benign apps tend to have significantly fewer triggers per app.

\vspace{0.2cm}

\subsubsection{Case Study}
\label{sec:goodware:case_studies}

This section presents an SHSO of a benign app.

\noindent
\textbf{Benign App.}
The app we consider in this case study is "no.apps.dn\-bnor".
\difuzer detected an SHSO in method \texttt{<bom.$\oint$: java.lang.\-String $\cdot\cdot\cdot\cdot$()>} which tests if the value of \texttt{Build.CPU\_ABI} or \texttt{Build.CPU\_ABI2} is equal to pre-defined values stored in a file.
In the case a match is found, it triggers the copy of a native code file into a second file.
The native code file name is in the form: "lib/" + \texttt{str} + "/lib" + \texttt{f9} + ".so".
The \texttt{str} variable represents a \texttt{CPU\_ABI} value and the \texttt{f9} variable represents a string to designate the file.
This file is then opened and eventually copied in the user data directory of the running app.

Although not malicious in this case, this behavior is suspicious, and \difuzer was able to reveal it.

\subsubsection{Malicious activities in \gp}
\label{sec:goodware:adware}
We now illustrate how \difuzer contributed to removing \appsRemovedFromGP apps whose behavior was potentially harmful to users (in the form of aggressive, unsolicited, and intrusive ads) in \gp.
Developers of such {\em adware apps} managed to evade classical checks performed in \gp.

During our manual analyses of benign apps, we stumbled upon an app with an SHSO flagged by \difuzer.
Our inspection of the code suggested that the SHSO is not a logic bomb per se since it does not trigger the malicious code.
However, during this manual analysis, we noticed that the app was apparently mainly designed to display advertising content aggressively.
To confirm our hypothesis, we downloaded the sample and executed it in an emulator.
First, we noticed poor app design, poor quality, and low content. Then in nearly every screen (i.e., \texttt{Activity} component), we received embedded ads and full-screen ads.
This behavior is characteristic of adware apps.
After verification, we found that the app was still in \gp with a relatively high number of downloads (a few thousands) but with negative comments.
In fact, the app pretended to provide users with a "walkthrough" version of an existing game to display a profusion of ads on each screen.

We then search in our analyzed apps if \difuzer detected similar SHSOs.
Eventually, \difuzer detected three apps with the exact same SHSO and the exact same service proposed to the user (walkthrough games).
We tested these apps to confirm they were adware.
They were also still in  \gp.

We then checked if similar "walkthrough games" were also still in \gp and not in our initial dataset.
Therefore, we searched for apps made by the same developers of the three previous apps detected by \difuzer.
We also searched for "walkthrough games" in  \gp and browsed the resulting apps.
We inspected the newly collected apps and confirmed they were adware apps.
Eventually, we identified \appsRemovedFromGP apps with the same adware behavior.

We contacted Google to report these \appsRemovedFromGP apps.
They were removed in less than two weeks from \gp.
We make available the samples in the project's repository.

\highlight{
\textbf{RQ3 answer:}
Our experiments show that SHSOs are present in benign apps and in widely-used libraries.
We have seen through real-world examples that \difuzer can reveal potentially harmful applications (PHA) and raise alarms concerning some apps' potential maliciousness.
Overall, \difuzer contributed to removing \appsRemovedFromGP adware apps from  \gp.
}

\section{Limitations and Threats to Validity}
\label{sec:limitations}

An essential step in our approach is the identification of SHSOs entry-points.
To do so, \difuzer relies on state-of-the-art tool \flowdroid~\cite{arzt2014flowdroid}.
Therefore, it carries the analysis limitations of \flowdroid, i.e., unsoundness regarding reflective calls~\cite{li2016droidra}, dynamic loading~\cite{xue2017auditing}, multi-threading~\cite{maiya2014race} and native calls~\cite{lin2011benchmark}.

Although our approach proved to be efficient to detect SHSOs and logic bombs, feature selection can impact the performances.
Indeed, feature engineering is a challenging task and can be prone to unsatisfactory selection since it does not capture \emph{everything}.

Besides, our approach is based on SHSO entry-points detection using taint analysis, which relies on sources and sinks methods.
Sinks are not an issue in our approach since they always represent \emph{if conditions}.
However, sources selection is at risk since they have been selected systematically, using heuristics and human intuitions.
Therefore, our list of sources might not be complete.

Although, we have implemented \ts by strictly following the description in the original paper, our implementation might not be exempt from errors.

In the absence of a-priori ground truth, some of our assessment activities rely on manual analysis based on our own expertise. While we follow a consistent process (e.g., we carefully verify the hidden behaviour implementation against the antivirus report), our conclusions remain affected by human subjectivity. Nevertheless, we mitigate the threat to validity by sharing all our artefacts to the research community for further exploitation and verification.
\section{Related work}
\label{sec:related_work}

\noindent
{\bf Logic bombs in general.} Hidden code triggered under specific conditions is a concern in many programming environments. The literature includes studies of the logic bomb phenomenon in programming prior to the Android era~\cite{chen2008towards, brumley2008automatically} and targeting the Windows platform for example.
Since then, various approaches have been proposed to tackle the challenging task of trigger-based behavior detection~\cite{shi2017detecting, jia2017findevasion, lindorfer2011detecting, kirat2014barecloud, balzarotti2010efficient}.
State-of-the-art techniques for the detection of trigger-based behaviour are varied and leverage fully-static analyses~\cite{papp2017towards, fratantonio2016triggerscope, zhao2020automatic}, dynamic analyses~\cite{zheng2012smartdroid}, hybrid analyses~\cite{brumley2008automatically, bello2018ares}, and machine-learning-based analyses~\cite{pan2017dark}.

\noindent
{\bf Trigger-based behavior detection for Android}
\difuzer combines static taint analysis and unsupervised machine learning techniques. Our closest related work is thus \hm~\cite{pan2017dark}, which relies on static analysis and automatic classification to detect \emph{HSOs}. Contrary to our work, however, \hm is not targeting suspicious HSOs and therefore does not focus on logic bombs.

Fratantonio et al.~\cite{fratantonio2016triggerscope} proposed \ts, an automated static-analysis tool that can detect logic bombs in Android apps.
\ts leverages a symbolic execution engine to model specific values (i.e., SMS-, time-, location-related variables).
\ts models conditions using \emph{predicate recovery}.
It combines symbolic execution results and path predicate recovery results to infer suspicious triggers.
Finally, potential suspicious triggers undergo a control dependency step to verify if it guards sensitive operations. Nevertheless, the whole approach relies on static analysis to check defined properties of suspiciousness. In contrast, \difuzer takes advantage of unsupervised learning to discover abnormal (hence suspicious) trigger-based behavior.

\noindent
{\bf Anomaly detection for security.} We note that the idea of using anomaly detection to detect malware has been presented in the Avdiienko et al.'s paper~\cite{7194594}.
Indeed, they present \textsc{MudFlow} that relies on anomaly detection to spot malware for which sensitive data flows deviate from benign data flows.
It proved to be efficient by detecting more than \num{86}\% malware.
While our approach is also based on anomaly detection to triage \emph{abnormal} triggers (i.e., suspicious sensitive behavior) that deviate from normality (i.e., normal triggers/conditions), the end goal of both approaches is different. 
Indeed, \textsc{MudFlow} addresses a binary classification problem to discriminate malware from goodware. 
In contrast, \difuzer addresses the problem of detecting and locating \emph{Suspicious Hidden Sensitive Operations} that are likely to be logic bombs in Android apps.

\noindent
{\bf Malicious behavior detection in Android apps.} 
Malware detection does not only focus on trigger-based malicious behavior.
Indeed, the Android security research community worked on tackling general security aspects~\cite{zhou2012dissecting, burguera2011crowdroid, lindorfer2014andrubis, tam2015copperdroid, mahindru2017dynamic}.
In the literature, numerous approaches have been proposed to detect Android hostile activities.
Among which, machine-learning techniques~\cite{sahs2012machine}, deep-learning techniques~\cite{mclaughlin2017deep}, static analyses through semantic-based detection~\cite{feng2014apposcopy}, privacy leaks detection~\cite{li2015iccta, arzt2014flowdroid, 9402001}, as well as dynamic analyses\cite{van2013dynamic, 10.1145/2592791.2592796, enck2014taintdroid}.
Each of these approaches tackles a particular aspect of Android security. Therefore, analysts could combine our approach with the aforementioned techniques to detect a wide variety of Android malicious behavior more efficiently.
\section{Conclusion}
\label{sec:conclusion}

We proposed \difuzer, a novel approach for detecting \emph{Suspicious Hidden Sensitive Operations} in Android apps. 
\difuzer combines bytecode instrumentation, static inter-procedural taint tracking, and anomaly detection for addressing the challenge of accurately spotting relevant SHSOs, which are likely logic bombs. 
After empirically showing that our prototype implementation can detect SHSOs with high precision (i.e., \difuzerPrecision\%) in less than 35 seconds per app, we assessed its capabilities to reveal logic bombs and demonstrate that up to \num{30}\% of detected SHSOs were logic bombs. We therefore improve over the performance of the current state of the art, notably \ts, which yields significantly more false positives, while detecting less logic bombs. Finally, we apply \difuzer on goodware to investigate potential SHSOs: \difuzer eventually contributed to removing \appsRemovedFromGP new adware apps from \gp. 
\section{Data Availability}
\label{sec:data_availability}

For the sake of Open Science, we provide to the community all the artifacts used in our study.
In particular, we make available the datasets used during our experimentations, the source code of our prototype, the executable used for our experiments, the annotated list of our manual analyses, and a dataset of logic bombs.

The project's repository including all artefacts (tool, datasets, etc.) is available at:

\begin{center}
    \url{https://github.com/Trustworthy-Software/Difuzer}
\end{center}
\section{Acknowledgment}
\label{sec:acknowledgment}
This work was partly supported 
(1) by the Luxembourg National Research Fund (FNR), under projects Reprocess C21/IS/16344458 the AFR grant 14596679, 
(2) by the SPARTA project, which has received funding from the European Union's Horizon 2020 research and innovation program under grant agreement No 830892, 
(3) by the Luxembourg Ministry of Foreign and European Affairs through their Digital4Development (D4D) portfolio under project LuxWAyS,
and (4) by the INTER Mobility project Sleepless@Seattle No 13999722.

\bibliographystyle{ACM-Reference-Format}
\bibliography{bib}

\end{document}